\documentstyle[12pt]{article}
\def\appendix{
\vskip 1cm
{\Large\bf Appendix}
\vskip 1cm
\par
\setcounter{equation}{0}
\def\theequation{A.\arabic{equation}}
}
\begin{document}
\thispagestyle{empty}
\renewcommand{\theequation}{\arabic{section}.\arabic{equation}}
\begin{flushright}
{\large \bf JINR preprint E2-93-181\\
Dubna, 1993}
\end{flushright}
\vskip0.5cm
\begin{center}
{\large \bf  Classical  dynamics of rigid string from Willmore
functional}\\[0.3cm]
{\large \bf A. L. Kholodenko$\dagger $ and V. V. Nesterenko$\ddagger $
\footnote {This work is supported in part by
Russian Foundation for Fundamental Research under project 93-02-3972.}}\\
\end{center}

$\dagger $375 H. L. Hunter Laboratories, Clemson University, Clemson,\\ SC
29634-1905
USA\\
E-mail address: string@clemson.clemson.edu \\

$\ddagger $Laboratory of Theoretical Physics, Joint Institute for Nuclear
Research,\\
Dubna SU-141980, RUSSIA \\
E-mail address: nestr@theor.jinrc.dubna.su
\vskip1cm
\centerline{\large \bf Abstract.}
     A new approach for investigating the  classical  dynamics
of the relativistic string model with rigidity is proposed. It
is based on the embedding of the string world surface into the
space of  a  constant  curvature.  It  is shown that the rigid
string in flat space-time is described by  the  Euler-Lagrange
equation  for  the  Willmore functional in a space-time of the
constant curvature  $K\,=\,-\,  \gamma/(2  \,  \alpha )$,   where
$\gamma$ and   $\alpha $  are   constants  in  front  of  the
Nambu-Goto term and the curvature term  in  the  rigid  string
action respectively.  For  simplicity the Euclidean version of
the rigid  string  in  the  three-dimensional  space-time   is
considered. The   Willmore  functional  (the  action  for  the
"Willmore string") is obtained by dropping the Nambu-Goto term
in the  Polyakov-Kleinert  action  for  the rigid string.  Such
"reduction" of the rigid string model would  be  useful,  for
example,   by applying some
results about the Numbu-Goto string dynamics in the de Sitter universe
to the rigid string model in the Minkowski space-time.
It allows also to use  numerous  mathematical  results
about the  Willmore  surfaces  in  the context of the physical
problem.
\newpage
\vspace*{1cm}
\section{Introduction}

        Polyakov-Kleinert (P-K)  rigid  string  model  [1, 2] has
been recently widely  considered  by  researchers  working  in
fields  ranging  from  particle  physics [3] and cosmology [4, 5]
through condensed matter [6] and biophysics [7].  For a recent
review, e.g. see [3, 8].

        In spite of the above interest of  various  groups  of
researchers  in  the rigid string model,  very little progress
has been achieved to date in our theoretical understanding  of
this  model.  Indeed,  unlike  the  Nambu-Goto (N-G) model [9]
which is  thoroughly  studied  both  classically  and  quantum
mechanically,  the  rigid  string model is not well understood
even at  the  classical  level.  If  the  traditional  methods
developed for N-G model are applied to rigid string,  then, at
the classical level,  the rigid string equations of motion are
nonlinear in any gauge.  The above nonlinearity precludes the
use of conventional quantization  methods  developed  for  N-G
model.  Accordingly,  for  the  rigid string model there is no
analog of the Virasoro algebra which allows to  determine  the
critical dimension of this model.

        In view of  situation  just  described,  most  of  our
knowledge   about   the   rigid  string  is  based  on  rather
inconclusive numerical simulation  results  which  employ  the
discretized   lattice   version   of  P-K  string  [10].  These
simulations typically involve  only  study  of  the  Euclidean
version  of  the  P-K  rigid  string.  In  case  of  Minkowski
space-time this model is having an  additional  ghost  states [2, 11].
Because  of  these  states,  the  rigid  string model has been
severely criticized recently [12].  The authors of Ref.\ 12  had
come  to the conclusion that "either fourth derivative kinetic
term most be quantized with  an  indefinite  norm...  or  with
energy unbound from below."

        The rigid string action is given by [1, 2]
\begin{equation}
A\,=\,\gamma \int \!\!\!\!\int dS\,+\, \alpha \int \!\!\!\! \int H^2dS
\end{equation}
where $\alpha $  and $\gamma $ are  some    constants.   In   (1.1),
integration  takes place over the string world surface $S$ which
has  extrinsic  mean  curvature  $H$.
For simplicity we confine ourselves to considering
the three-dimensional space-time and to the Euclidean version of this model.
In order to rid of the boundary conditions we shall treat closed string
world surfaces which are encountered, for example, in string vacuum functional.

The equations of motion for action (1.1) written in terms of the
string coordinates are very complicated. They stand for a system of nonlinear
partial differential equations of the fourth order [13].
Except for one model example [14],
nothing is known about solutions of such system of  equations.
However, varying the functional (1.1) one can arrive at quite simple
equations relating basic geometrical invariants of the string world surface,
its Gauss curvature  $K$ and mean curvature $H$.
We shall be working mainly  with
equations of such type.

        The main result of  our  work  can  be  formulated  as
follows.  We  are going to demonstrate that the effects of N-G
term in the total action $A$ (the first term in equation  (1.1))
could  be  accounted  for  by considering the truncated action
which contains only the second term  in  (1.1)  provided  that
this  term is considered not in flat but in curved space-time.
Thus truncated action is known in the literature  as  Willmore
functional [15] and,  whence, we shall call string model, based
on such truncated action, as Willmore string. Reduction of the
variational  problem for the rigid string action $A$ to that for
Willmore string is advantageous for number of reasons.

        First, because  the  action  (1.1)  is two dimensional
extension   of   the   action   used   for   particles    with
curvature-dependent  action  [16, 17],  it is logically natural to
search for methods which extend those developed for  particles
to that used for strings.
        In Ref.\ 18 the non-relativistic quantum  mechanics  of
point-like  particles is formulated on the surface of 3-sphere
$S^3$ embedded in $R^4$.  Such reformulation leads to the  emergence
of   spin   for   initially  spinless  particle.  Because  the
statistical mechanics of spinless particles  is  the  same  as
fully  flexible  polymers,  the  presence  of  spin for such
particles leads to the effective rigidification  of  initially
flexible  polymers  [19].  Alternatively,  such  rigidification
could be  achieved  if  the  above  "particle"  moves  in  the
presence  of  (in  general)  (non)abelian monopole gauge field~[20].
  In this case the rigidification mechanism lies  in  the
replacement  of  ordinary  derivatives  in "flexible particle"
Hamiltonian by covariant ones causing our particle to move  in
the  effective gauge (gravity) field.  The present work can be
viewed as extension of the above ideas  to  the  case  of  two
dimensional objects, e.g., rigid strings.

        Second, in modern geometry there had been  accumulated
considerable   amount   of   results   related   to   Willmore
functionals, e.g. see Ref. [21, 22], so that our understanding
of rigid strings (at least at the classical level), in view of
results  of  our  work,  will  depend  to  large   extent   on
appropriate  interpretation  and  utilization  of  the already
accumulated knowledge.

        The layout of the paper is as follows.
In Section 2 we provide auxiliary facts from classical
differential geometry  of  surfaces  in  order  to  arrive  at
Willmore's  "equation of motion" connecting Gaussian curvature
$K$ with mean curvature  $H$  in  flat  embedding  space-time.  In
Section  3  we  extend  the above results to the case when the
embedding space is the space of  constant  curvature.  In  the
context  of quantum field theories the problem of embedding of
the corresponding field-theoretic model into space of constant
negative   curvature   was  recently  considered  in  [23]  in
connection with improved infrared regularization  of  QCD.  In
our  case  we  study the embedding with different purpose.
By doing so we are hoping to apply some results about the Nambu-Goto string
dynamics
in the  de Sitter universe  to the rigid string model in the Minkowski
space-time.
        Finally, in Section 4 (Conclusion) we provide a brief  summary  and
discussion  of  possible  future  applications of the obtained
results. In Appendix a more simple
one-dimensional version of our problem is considered. Instead of surfaces
 in action (1.1)  we are dealing here with  curves.

\section{Normal variations of the surfaces}
\setcounter{equation}0
For the completeness we give here the basic equations from the
classical differential geometry of the surfaces that will
be required in the following [24].

Let $x^\mu(u^1,\,u^2), \quad \mu =1,\,2,\,3$ be a parametric representation of
the
surface $M$ in the three dimensional Euclidean space $E^3$ and $n^\mu$ is a
unit normal to
the surface. Intrinsic  differential geometry of the surface is defined by the
induced metric or the first quadratic differential form of the surface
\begin{equation}
g_{i,j}(u^1,\,u^2)\,=\,x^\mu_{,i}x^\mu _{,j},\quad x_{,i}^\mu \,
\equiv \,\frac{\partial x^\mu (u^1,\,u^2)}{\partial u^i},\quad i,\,j\,=\,1,\,2.
\end{equation}
The central point of the surface theory is the derivation equations of Gauss
\begin{equation}
x^\mu_{,ij}\,=\, \Gamma ^k_{ij}\,x^\mu _{,k}\,+\,b_{ij}n^\mu
\end{equation}
and Weingarten
\begin{equation}
n^\mu _{,i}\,=\,-\,b_{ij}g^{jk}x^\mu _{,k}.
\end{equation}

Here $\Gamma _{ij}^k$ are the Christoffel symbols for the metric $g_{ij }$
[24],
$g^{ij}$ is an inverse matrix to $g_{ij},\quad b_{ij}$ are the coefficients
of the second quadratic form of the surface that determines its external
curvature $(b_{ij}\,=\,b_{ji})$.

Equations (2.2) and (2.3) describe the motion of the basis
$\{x^\mu _{,1}\,x^\mu _{,2},\,n^\mu \}$ along the surface. The compatibility
conditions of
these linear equations are given by the  Gauss equation
\begin{equation}
R_{ijkl}\,=\,b_{ik}b_{jl}\,-\, b_{il}b_{jk}
\end{equation}
and by the Codazzi equations
\begin{equation}
b_{ij;k}\,-\, b_{ik;j}\,=\,0,\quad i,\,j,\,k\,=\,1,\,2.
\end{equation}
The semicolon means the covariant differentiation
with respect to the metric tensor $g_{ij}$ in (2.1) and $R_{ijkl}$ is the
Riemann curvature tensor
[24].

When the equations (2.4) and (2.5) are satisfied by given tensors
$g_{ij}$ and $b_{ij}$ then the derivation equations (2.2) and (2.3) can be
integrated
and their  corresponding  solution $x^\mu (u^1,\,u^2)$ determines the surface
up to its motion
in $E^3$ as a whole.

The important geometrical invariants of the surface
are its Gaussian curvature
\begin{equation}
K\,=\,-R/2, \quad R\,=\,g^{il}g^{jk}R_{ijkl}
\end{equation}
and its mean curvature
\begin{equation}
H\,=\,\frac{1}{2}\,b_{ij}g^{ij}\,=\,\frac{1}{2}\,b_i^i\,{.}
\end{equation}

For physical applications dealing with
closed surfaces it is sufficient to consider {\t normal
variations} of the surface that are defined  as follows.
For a given surface $M$ with a position vector $x^\mu (u^1,\,u^2)$  we
form the surface $\bar M$ {\it parallel} to $M$  putting
\begin{equation}
\bar x ^\mu\,=\,x^\mu \,+\,t\,f\,n^\mu,\quad
-\varepsilon < t < \varepsilon ,
\end{equation}
where $f(u^1,\,u^2)$ is a sufficiently smooth function
given on $M$. We denote by $\delta $ the operator $\partial /\partial
t\,|_{t=0}$.
Thus $\delta x^ \mu = fn^\mu$. For simplicity, we shall omit the bar at the
argument
of $\delta $.

 From the definition (2.8) we obtain
\begin{eqnarray}
\delta \,x^\mu _{,i} &=&f_{,i}n^\mu \,+\, f \,n^\mu _{,i}, \\
\delta \,x^\mu _{,ij} &=&f_{,ij}n^\mu \,+\,f_{,i}n^\mu _{,j}\,
+\,f_{,j}n^\mu _{,i}\,+fn^\mu _{,ij}.
\end{eqnarray}
The variation of the metric tensor
(2.1) is given by
\begin{equation}
\delta g_{ij}\,=\,\delta x^\mu _{,i}\,x^\mu _{,j}\,+\,x^\mu _{,i}\delta x^\mu
_{,j}\,=\,
f(n^\mu _{,i} x^\mu_{,j}\,+\,n^\mu _{,j}x^\mu _{,i}).
\end{equation}
By making use of the Weingarten derivation
equation (2.3) the last  equation can be rewritten as
\begin{equation}
\delta g_{ij}\,=\,-2fb_{ij}.
\end{equation}
By varying the definition
$$
g_{ij}g^{jk}\,=\,\delta_i^k
$$
we have
$$
\delta g_{ij}g^{jk}\,+\,g_{ij}\delta g^{jk}\,=\,0.
$$
Whence,
\begin{equation}
\delta g^{lk}\,=\,-g^{li}g^{jk}\delta g_{ij}\,=\,2f b^{lk}.
\end{equation}

Denoting, as usual, by $g$ a determinant of the
metric tensor, $g=\det (g_{ij})$, we can write
\begin{equation}
\delta \sqrt{g}  =  \left ( \frac{\partial }{\partial x^\mu_{,i}}
\sqrt {g} \right )\delta x^ \mu_{,i}\,=\, \sqrt {g}g^{im}
x^\mu _{,m}n^\mu _{,i}f\,=\,-2 \sqrt {g}H f.
\end{equation}
 From (2.2) it follows that
\begin{equation}
b_{ij}\,=\,n^\mu x ^\mu _{,ij}.
\end{equation}
Hence
\begin{eqnarray}
\delta b_{ij}&=& \delta n^\mu x ^\mu _{,ij}\,+\,n^\mu \delta
x^\mu _{,ij}\,=\nonumber \\
&=& \Gamma^k_{ij}\delta n^\mu x^\mu _{,k}\,+\,b_{ij}\delta n^\mu\,n^\mu
\,+\,n^\mu \delta \,x^\mu _{,ij}\,{.}
\end{eqnarray}
By varying the equalities following from the definition of the
normal $n^\mu $
\begin{equation}
n^\mu n_\mu\,=\,1, \quad n^\mu x^\mu _{,i}\,=\,0
\end{equation}
we get
\begin{equation}
\delta n^\mu n^\mu \,=\,0,
\end{equation}
\begin{equation}
\delta n^\mu x^\mu_{,i}\,=\,-n^\mu \delta x^\mu _{,i}\,=\,-f_{,i}.
\end{equation}
In addition, one can write
\begin{equation}
n^\mu \,n^\mu _{,ij}\,=\,-n^\mu_{,i}\,n^\mu _{,j}\,=\,- b_{ik}\,b_{jl}\,g^{kl}.
\end{equation}
Finally the variation of the second quadratic form  is given by
\begin{equation}
\delta b_{ij}\,=\,f_{;ij}\,-\,f\,b_{ik}\,b_{jl}g^{kl}.
\end{equation}
Now we can calculate the variation of $H^2$:
$$
\delta H^2\,=\,H\,\delta \,(b_{ij}\,g^{ij})\,=\,H\,g^{ij}\,\delta b_{ij}\,+\,
H\,b_{ij}\,\delta g^{ij}\,=
$$
\begin{equation}
=\,H\,(\Delta f\,+\,f\,b_k^ib_i^k),
\end{equation}
where $\Delta $ is the Laplace-Beltrami operator given on the surface $M$. From
the
Gauss equation (2.4) it follows that
\begin{equation}
R\,=\,b^i_kb_i^k\,-\,b_i^ib_k^k\,=\,b_k^ib_i^k\,-\,4\,H^2.
\end{equation}
Thus the variation $\delta H^2$  acquires the final form
\begin{equation}
\delta\,H^2\,=\,H\,[\Delta f\,+\,f\,(R\,+\,4\,H^2) ].
\end{equation}

 Now we can derive the Euler-Lagrange equation following from the vanishing of
the
normal variation of the rigid string action (1.1)
$$
\delta A \,=\,\delta \int \!\!\! \int (\gamma \,+\, \alpha \,H^2)\,dS\,=
\,0, \quad dS\,=\,\sqrt {g}\,du^1\,du^2.
$$
By making use of eqs. (2.14) and (2.24) we obtain
\begin{equation}
\delta \,A\,= \int \!\!\!\int dS\,\left \{ [-2\,\gamma \,H\,+\,\alpha \,(2
H^3\,+\,H\,R) ]\,f
\,+\,\alpha \,H \Delta f
 \right \} \,=\,0.
\end{equation}
On the closed surfaces the Laplace-Beltrami operator $\Delta $ is  selfadjoint
operator [25]
$$
\int dS\,\varphi \Delta f \,=\,\int dS\,f \Delta\varphi,
$$
therefore the variation $\delta A$ in (2.25) can
be rewritten as follows
$$
\delta \,A \,=\int dS\,[-2 \gamma H\,+\,\alpha \,(\,\Delta H\,+\,
2H^3\,+\,H\,R\,)]\,f\,=\,0.
$$
Due to the arbitrariness of the function $f(u^1,\,u^2)$ we arrive at the
equation of motion
\begin{equation}
-2\,\gamma \,H\,+\,\alpha \,(\,\,\Delta \,+\,2H^3\,+\,H\,R H)\,=\,0.
\end{equation}

We gave here quite  detailed derivation of eq.~(2.26) that is rather well known
in literature [15, 26] in the case of a Euclidean  ambient
space $E^3$. We shall use the methods, just described,
in the next section for deriving
the equation on the geometrical invariants  $H$ and $R$ when the string
world surface is placed in a space-time of a constant curvature $S^3$.
\section{Willmore surfaces in a space of a constant curvature}
\setcounter{equation}0
Here we show that the equation of motion (2.26) can be derived by considering
the Willmore surfaces in a space of a constant  curvature $S^3$. The Willmore
surfaces are extremals of the
Willmore functional
\begin{equation}
W\,=\,\int\!\!\!\int dS\,H^2.
\end{equation}

By making use of the Weierstra\ss \ coordinates $z^\alpha ,\quad \alpha
=1,...,4$   [24, 27] the three-dimensional
sphere $S^3$ with radius $a$ can be represented as a hypersurface in  the four
dimensional
Euclidean space $E^4$
\begin{equation}
\sum_{\alpha =1}^{4}z^\alpha \,z^\alpha \,=\,a^2.
\end{equation}

Let $z^\alpha (u^1,\,u^2),\;\;\alpha =1,\,2,\,3,\,4$ is a parametric
representation of the surface $M$ embedded into $S^3$ in term of the
Weierstra\ss \ coordinates obeying (3.2). The natural unit normal
to this surface in $E^4$ is $z^\alpha (u^1,\,u^2)$ and let $n^\alpha $ be the
second unit
normal to this surface
\begin{equation}
\sum_{\alpha =1}^{4} n^\alpha \,n^\alpha \,=\,1,\quad\sum_{\alpha =1}^{4}
n^\alpha \,z^\alpha \,=\,0,\quad \sum_{\alpha =1}^{4}
n ^\alpha \,z^\alpha _{,i}\,=0.
\end{equation}

The important advantage of the Weierstra\ss \ coordinates in
 the problem  under consideration is the following. The basic
equations for the surface embedded into  $S^3$  are very simple,
they are almost the same as in the Euclidean ambient space. For the metric
tensor
on $M$ we have now
\begin{equation}
g_{ij}\,=\,\sum_{\alpha  =1}^4{}z^\alpha _{,i}\,z^\alpha_{,j}\,=\,-\sum_{\alpha
=1}^4{}z^\alpha_{,ij}\,z^\alpha .
\end{equation}
The derivation equations (2.2) and (2.3) become [24]
\begin{equation}
z^\alpha _{,ij}\,=\,\Gamma ^k_{ij}\,z^\alpha _{,k}\,+\,b_{ij}\,n^\alpha
\,-\,\frac{g_{ij}}{a^2}\,z^\alpha ,
\end{equation}
\begin{equation}
n^\alpha _{,i}\,=\,-\,b_{ij}\,g^{jk}z^\alpha _{,k}.
\end{equation}
The Gauss equation (2.4) now reads
\begin{equation}
R_{ijkl}\,=\,b_{ik}\,b_{jl}\,-\,b_{il}\,b_{jk}\,+\,\frac{1}{a^2}\,
(g_{ik}\,g_{jl}\,-\,g_{il}\,g_{jk}).
\end{equation}
The Codazzi equations (2.5) keep their form.

The normal variation in terms of the Weierstra\ss \ coordinates is defined
as follows
\begin{eqnarray}
\bar z^\alpha &=&z^\alpha \,+\,t\,f\,n^\alpha ,\quad -\varepsilon
< t<\varepsilon , \nonumber \\
\delta z ^\alpha &=& f\, n^\alpha, \nonumber \\
\delta\, z^\alpha _{,i}& =& f_{,i}\,n^\alpha \,+\, f\,n^\alpha _{,i}, \\
\delta \,z^\alpha _{,ij}&=&f_{,ij}\,n^\alpha \,+\,f_{,i}n^\alpha _{,j}\,+\,
f_{,j}n^\alpha_{,i}\,+\,f\,n^\alpha _{,ij}.\nonumber
\end{eqnarray}
By making use of (3.2) and (3.3)  one can easily convinced that such a
variation does
not take out from $S^3$.

For variation of the metric tensor (3.4) we have obviously the same equations
(2.12) -- (2.14).
 From (3.5) it follows that the coefficients of the
second fundamental form $b_{ij}$ are defined by
\begin{equation}
b_{ij}\,=\,\sum_{\alpha =1}^{4}n^\alpha \,z^\alpha _{,ij}.
\end{equation}
Therefore,
 \begin{eqnarray}
\delta \,b_{ij} &=& \delta n^\alpha\,z^\alpha _{,ij}\,+\,n^\alpha\, \delta
z^\alpha _{,ij}=
\nonumber \\
&=&\delta n^\alpha\,\left (  \Gamma ^k_{ij}z^\alpha _{,k}\,+\,b_{ij}n^\alpha
\,-\,
\frac{g_{ij}}{a^2}\,z^\alpha \right )\,+\,n^\alpha \delta z^ \alpha_{,ij}.
\end{eqnarray}
For simplicity we omit here and in the following the sign of summation with
respect to repeated indices.

 From (3.2), (3.3) and (3.8) it follows that
$$
\delta n^\alpha \,n^\alpha \,=\,0, \quad n^\alpha_{,i}\,n^\alpha \,=\,0, \quad
\delta n^\alpha z^\alpha \,=\,- n^\alpha \delta z^ \alpha \,=\,-f,
$$
\begin{equation}
\delta n^\alpha z^\alpha_{,i}\,=\,-n^\alpha \delta z^\alpha _{,i}\,=\,-f_{,i},
\end{equation}
$$
n^\alpha \delta z^\alpha_{,ij}\,=\,f_{,ij}\,+\,fn^\alpha n^\alpha_{,ij}\,=\,
f_{,ij} \,-\,fn^\alpha _{,i}n^\alpha_{,j}.
$$
Now equation (3.10) becomes
\begin{equation}
\delta b_{ij}\,=\,f_{,ij}\,-\,\Gamma
^k_{ij}f_{,k}\,+\,\frac{g_{ij}}{a^2}\,f\,-\,fn^\alpha_{,i}n^\alpha _{,j}.
\end{equation}
With allowance of (3.6) we obtain
\begin{eqnarray}
\delta b_{ij}&=&f_{;ij}\,+\,\frac{g_{ij}}{a^2}\,f\,
-\,f\,b^k_ib^l_j\,z^\alpha _{,k}z^\alpha_{,l}\,=\, \nonumber \\
&=&f_{;ij}\,+\,f\,\left (\, \frac{g_{ij}}{a^2}\,-\,b^k_ib_{jk}\,\right )
\end{eqnarray}
By making use of the Gauss equation (3.7) we deduce now instead of (2.23)
\begin{equation}
R\,=\,b_k^i\,b_i^k\,-\,4\,H^2\,-\,\frac{2}{a^2}.
\end{equation}
Whence
\begin{equation}
\delta \,H^2\,=\,H\,\left [\Delta f\,+\,f\,\left (R\,+\,4 H^2\,
+\,\frac{4}{a^2}\right )\right ].
\end{equation}
Taking into account that
\begin{equation}
\delta \,dS\,=\,-\,2H\,f\,dS
\end{equation}
we can write
\begin{equation}
\delta \,W\,=\,\int \!\!\!\int dS\left [ \Delta H \,+\,2H^3\,+\,H\, \left (
R\,+\,\frac{4}{a^2}
\right )\,\right ]\,f.
\end{equation}
Therefore the equation of motion for the Willmore string in $S^3$ is
\begin{equation}
\Delta H \,+\,2H^3\,+\,H\, \left ( R\,+\,\frac{4}{a^2} \right )\,=\,0\,{.}
\end{equation}
Thus it has the same form as (2.26) if we put
\begin{equation}
\frac{\gamma }{\alpha}\,=\,-\,\frac{2}{a^2}\,{.}
\end{equation}
This result is in a complete agreement with an analogous relation
in the one-dimensional version
of the problem under consideration (see eq. (A.5)  in Appendix and
take into account that the
sectional curvature $G$ in the case of the sphere (3.2) is equal to $1/a^2$).

We would like to note here the following. In spite of
the Willmore surfaces in  spaces with curvature have been
considered in a number of mathematical papers [28 -- 30] nevertheless a
simple derivation of eq. (3.18)
in a  correct form are given here  actually for the first time.
Indeed, if we try to apply the final equation (3.13) from  Ref. [28] for
embedding two-dimensional Willmore surface
into the $S^3$ we obtain instead of the second term  $4/a^2$ in brackets in
(3.18)
the wrong expression $3/a^2$. In paper [29] a more general functional as
compared  with (3.1) has been considered. For closed surfaces it reads
\begin{equation}
W_1\,=\,\int\!\!\!\int (H^2\,+\,\tilde K)\,dS,
\end{equation}
where $\tilde K$ is a constant sectional curvature of the ambient space. Taking
into account that the
subtraction from $W_1$ of the functional
\begin{equation}
-\,\tilde K \int\!\!\!\int dS
\end{equation}
results in the additional term in the Euler-Lagrange equation
$+2\, \tilde K\,H$ one obtains from eq.~(14) of paper [29] our result (3.18).
In the same time, the final equations (5.43) and (5.46) in paper [30] cannot be
compared directly with our eq. (3.18).
In order to do this they should be combined with the Gauss equation.

The result obtained here (eq.~(3.18)) can be generalized
directly in the following two ways. At first, if we consider the Willmore
surfaces not
in the $S^3$ but in the three dimensional manifold of a constant negative
curvature (the pseudosphere with imaginary radius $ia$) then eq. (3.18) becomes
\begin{equation}
\Delta H \,+\,2H^3\,+\,H\, \left ( R\,-\,\frac{4}{a^2} \right )\,=\,0\,{.}
\end{equation}
Instead of (3.19) we have in this case
\begin{equation}
\frac{\gamma }{\alpha}\,=\,\frac{2}{a^2}\,{.}
\end{equation}
Secondly, we can generalize our result to the $d$-dimensional hypersurfaces
in the $S^{d+1}$  determined by a functional
\begin{equation}
W_2\,=\,\int \!\!\!\int H^m\,dS, \quad H\,=\, \frac{1}{d}\,g^{ij}b_{ij},\quad
i,j\,=\,1,2,\ldots , d,\quad m\,>\,0.
\end{equation}
In this case equation (3.18) becomes
\begin{equation}
\Delta H^{m-1}\,+\,d^2 \left ( 1\,-\,\frac{1}{m} \right )\,H^{m+1}+\,H^{m-1}
\left (
R\,+\, \frac{d^2}{a^2} \right )\,
=\,0 .
\end{equation}

When the Willmore $d$-dimensional surface is embedded into the
sphere $S^{d+n+1}$ with $n\;>\,1$ then one arrives at the $n$ equations
relating
internal and external characteristics of this surface in a complete analogy
with
one-dimensional case (see Appendix).
\section{Conclusion}
     Recently the string model based only on the  second  term
in eq.~(1.1) has been considered in paper [31].
 It was called as a spontaneous string
alluding to  the  fact  that  in this case the Nambu-Goto term
with nonzero string tension can  be  generated  spontaneously
due to  the quantum fluctuations.  We have proposed here other
classical scenario for this situation.

As a special solution to eq. (3.18) we can consider the minimal surfaces in
$S^3$
with $H=0$. There are some new results about these solutions obtained
under consideration of the usual Nambu-Goto string
in the space-time of a constant curvature [21, 22, 27, 32]. In particular,
authors of Ref. [32]
arrive at the conclusion that the dynamics of the Nambu-Goto string
 in the de Sitter space-time should be unstable. This instability turns out to
be a direct
consequence of the unboundness of the Hamiltonian of the Sinh-Gordon equation
that describes minimal surfaces in the three-dimensional  de Sitter universe
[27].
The relationship between the rigid string in flat space-time and the Willmore
string in $S^3$ enables us to
argue that the same instability should take place in the rigid string model in
flat space-time. In mathematics another relation
between minimal surfaces in $S^3$ and the Willmore surfaces in $R^3$ is known
[15, 21].
Applying a stereographic projection to minimal surfaces in $S^3$ one obtains
Willmore
surfaces in $R^3$. Whence, we can conclude that the Willmore string in $R^3$ is
unstable also.

 And the final note concerns a modified version of
the Willmore functional in $S^3$.
 From the physical point of view it is  desirable to preserve the conformal
invariance
of this functional in the case of ambient space with a nonzero curvature too.
To this end one has to use a modified form of it given in (3.20).
\appendix
We consider here more simple version of our problem, i.e. one dimensional
version of it. Let us introduce two functionals defined on the curves $x^\mu
(s)$:
\begin{eqnarray}
F_1&=& m\int ds \,+\,\alpha \int k^2 ds, \\
F_2&=& \int k^2 ds,
\end{eqnarray}
where $k$ is a curvature of the curve.
First functional we shall consider in the Euclidean space $E^n$ and the second
one
in the $n$-dimensional manifold
of constant sectional curvature $G$.
When $n=2$ it has been shown in the book [33] that the Euler-Lagrange equations
are identical for these two
problems. This result can be generalized easily to  arbitrary $n$.
By making use   the results of papers [34]
we can write the corresponding equations of motions. In the first case we have
$$
2\,k_{ss}\,+\,k^3 -2k \tau ^2 -\frac{m}{\alpha }\,k\,=\,0,
$$
\begin{equation}
k^2 \tau \,=\, \mbox { const},
\end{equation}
$$
k_i \,=\,0,\quad i\not=1,\,2.
$$
Here subscribe $s$ means differentiation with respect to the curve length,
 $\tau $ is the torsion of the curve and $k_i, \quad i=3, 4, \ldots , d-1$ are
the
higher curvatures of the curve.
For the functional $F_2$ the Euler-Lagrange  equations read
$$
2\,k_{ss}\,+\,k^3 -2k \tau ^2 \,+\, 2k G\,=\,0,
$$
\begin{equation}
k^2 \tau \,=\, \mbox{const},
\end{equation}
$$
k_i \,=\,0,\quad i\not=1,\,2.
$$

Thus we get identical systems if we put
\begin{equation}
\frac{m}{\alpha }\,=\,-\,2 G.
\end{equation}

\vskip1cm
\centerline{\large \bf     References }
\vskip0.5cm
\begin{enumerate}
\item Polyakov, A.\ M.: Nucl. Phys. {\bf B286}, 406 (1986)
\item Kleinert, H.: Phys. Lett. {\bf B174}, 335 (1986)
\item Germ\'an, G.: Mod. Phys, Lett. {\bf A20}, 1815 (1991)
\item  Gregory, R.: Phys. Lett. {\bf B206}, 199 (1988)
\item Maeda, K., Turok, N.: Phys. Lett. {\bf B202}, 376 (1988)
\item Keller, J., Merchant, J.: J. Stat. Phys. {\bf 63}, 1039 (1991)
\item  Jenkins, J: J.\ Math. Biology {\bf 4}, 149 (1977)
\item Zhang, X., Zhong-Can, O-Y.: Mod. Phys. Lett. {\bf B6}, 917 (1992)
\item Barbashov B.\ M., Nesterenko V.\ V.: Introduction to the relativistic
string theory. Singapore: World Scientific 1990
\item Baillie,  C.,  Johnson,  D.:  Phys. Lett. {\bf B295}, 249 (1992);
Phys. Rev. {\bf D46}, 4761 (1992)
\item Nesterenko, V.\ V.\ and Nguyen Suan Han: Int. J. Mod. Phys. {\bf 3A},
2315 (1988)
\item  Polchinski, J., Yang, Z.: Phys. Rev. {\bf D46}, 3667 (1992)
\item Arod\'z, H., Sitarz, A. and W\c{e}grzyn, P.: Acta Phys.\ Polonica {\bf
B22},
495 (1991)
\item Curtright, T.\ L., Ghandour, G.\ I., Thorn C.\ B., and Zachos C.\ K.:
Phys. Rev.
Lett. {\bf 57}, 799 (1986); Curtright, T.\ L., Ghandour, G.\
I., and Zachos C.\ K.: Phys. Rev. {\bf
D34}, 3811 (1986)
\item Willmore, T. J.: Total Curvature in Riemannian Geometry. Chichester~:
Ellis Horwood 1982
\item Nesterenko V.\ V.: J.\ Phys.\ A: Math.\ Gen.\ {\bf 22}, 1673 (1989);
Class. Quantum Grav. {\bf 9}, 1101 (1992);
J.\ Math.\ Phys. {\bf 32}, 3315 (1991)
\item Dereli, T., Hartley, D.\ H., \"Onder, M., and Tucker, R.\ W.: Phys.\
Lett.
{\bf B252}, 601 (1990)
\item  Ohnuki, Y., Kitakado, S.: Mod. Phys. Lett. A7, 2477 (1992)
\item Kholodenko, A.: Ann. Phys. 202, 186 (1990)
\item  Jaroszewich,  T.,  Kurzepa, P.: Annals of Physics 213, 135
(1992)
\item Bobenko, A.: Math. Ann. 290, 209 (1991)
\item Walter, R.: Manuscr. Math. 63, 343 (1989)
\item Callan, C., Wilczek, F.: Nucl. Phys. B340, 366 (1990)
\item Eisenhart, L.\ P.: Riemannian Geometry. Princeton, NJ: Princeton
University Press, 1964
\item Lichnerowicz, A.: The\'eorie Globale des Connexious et des
groups D'Holonomie. Roma Edizioni Cremonese, 1955
\item Chen, B.: Total Mean Curvature and Submaniflods of Finite
Type. Singapore: World Scientific, 1990
\item Barbashov, B.\ M. and Nesterenko, V.\ V.:  Commun. Math. Phys. {\bf 78},
499 (1981)
\item Willmore, T.\ J. and Jhaveri, C.: The Quarterly Journal of Mathematics,
{\bf 23},
319 (1972)
\item Weiner, J.\ L.: Indiana University Mathematics Journal {\bf 27}, 19
(1978)
\item Hartley, D.\ H. and Tucker, R.\ W.: In "Geometry of
low-dimensional manifolds" v. 1, p. 207. Cambridge: Cambridge University Press,
1990
\item Kleinert, H.: Phys. Lett. {\bf B189}, 187 (1987)
\item De Vega, H.\ and Sanchez, N.: Phys. Rev. {\bf D47} 3394 (1993)
\item Griffith, P. A.: Exterior differential systems and the calculus
of variations. Birkh\"auser, 1983
\item Langer, J. and Singer,D.\ A.: J. Lond. Math. Soc. {\bf 30}, 512 (1984);
J. Diff. Geom. {\bf 20}, 1  (1984)
\end{enumerate}

\vfill
\begin{center}
Received by Publishing Department \\
on May 21, 1993
\end{center}

\end{document}